\documentstyle[aps,prl,twocolumn]{revtex}

\newcommand{\tdot}[1]{\stackrel{\textstyle ...}{#1}}
\newcommand{\Ld}[1]{ { {2e^2}\over{3c^3} } #1}
\newcommand{\uni}[1]{ {\bf \hat{ #1}} }
\newcommand{\Le}[1]{ { {2e}\over{3c^3} } #1}

\begin{document}

\draft

\title{Electrodynamics of helium with retardation and diffusion in the neighborhood of Wannier orbits...}
\author{Jayme De Luca}
\address{Instituto de Qu\'{\i}mica de Araraquara-UNESP\\ Caixa Postal 355, Araraquara, S\~ao Paulo, 14801-970}

\maketitle
\date{\today}
\twocolumn[

\widetext
\vspace*{-1.0truecm}

\begin{abstract}
\begin{center}
\parbox{14cm}{We show that an extra constant of motion with an analytic form can exist in the neighborhood of some discrete Wannier orbits of helium, when one includes retardation and self-interaction effects. The energies of these discrete stable Wannier orbits are in the correct atomic magnitude. The highest frequency in the stable manifold of one such orbit agrees with the highest frequency sharp line of parahelium to within $2\%$. The generic term of the frequency in the stable manifold to higher orbits is also in agreement with the asymptotic form of quantum mechanics for helium.}
\end{center}
\end{abstract}

\pacs{
\hspace{1.9cm}
} 
]

\narrowtext

In this letter we explore some surprising consequences of the retardation effects of Maxwell's electrodynamics to atomic physics. We show that electrodynamics with retardation prescribes a discrete set of stable \hbox{Wannier} orbits for the helium atom. A Wannier orbit is one where the two electrons are in the same
circular orbit and in phase opposition, that is, along a diameter\cite{Poirier}. 
\par
The linearized dynamics about a Wannier orbit has a one-dimensional stable direction, a one-dimensional unstable direction and ten neutrally stable oscillatory modes\cite{Poirier}. For an infinitely massive $\alpha$-particle, helium is a six-degree of freedom Hamiltonian system with only four independent constants of motion, (namely, the energy and the three components of the total angular momentum). In the neighborhood of Wannier orbits, even if one restricts to the center manifold\cite{Gucken}, one still has a five-degree of freedom system. Because there are only four constants, the dynamics in the neighborhood of a generic Wannier orbit can diffuse away to large distances of the orbit. Here we show that an extra complex constant can exist in the neighborhood of a discrete set of Wannier orbits, when one includes retardation effects.   
\par
To include the retardation, we use the work of Page (1918)\cite{Page} on the expansion of the Li\'enard-Wiechert formula\cite{Jackson}. This formula is complicated because of the retardation constraint, and one way to convert it into a useful differential equation is to develop the constraint in a Taylor series. This was done by Page up to the fifth order in 1918\cite{Page}.   
Truncated to second order in $(v/c)$, this retarded interaction is described by the Darwin Lagrangian\cite{Jackson}, which introduces the first retardation correction to the Coulomb interaction. 
\par
The Darwin Lagrangian is rotationally invariant, which generates an angular momentum-like constant of the motion according to Noether's theorem\cite{Arnold}. This constant is the angular momentum of the electrons plus the small correction of the angular momentum in the electromagnetic field.  There is also an energy-like constant (because the Darwin Lagrangian is time independent).  
According to a recent result, the existence of a constant of motion analytic in the neighborhood of an equilibrium point requires the eigenvalues of the linearized dynamics to satisfy some resonance condition\cite{Furta}. Since the eigenvalues in the presence of retardation depend on the particular orbit, an extra complex constant is possible for some discrete orbits, which is a genuine signature of the nonlinear dynamics with retardation. We stress that it is absolutely essential to include the retardation to unfold a degeneracy of the Coulomb dynamics, and not to attain some better precision. Last, we consider only the dynamics of spinless pointlike charges, and issues like quantum mechanical spin are not discussed.
\par 
In this work we assume that the self-interaction effects are described by the Lorentz-Dirac
equation (LDE) with a renormalized mass\cite{Jackson,Coleman}.
The LDE equation for an electron of charge $-e$ can be written in the convenient, non-covariant form as
\begin{eqnarray} 
{{d}\over{dt}}(\gamma M_e \dot {\bf x}_e) & = & {\bf F}_{rad} + {\bf F}_{ext} , 
\label{eq:ld}
\end{eqnarray}
where $\dot {\bf x}_e$ is the electron's velocity, $M_e$ the renormalized electronic mass\cite{Coleman} and $\gamma\equiv 1/\sqrt{1-(|\dot {\bf x}_e|/c)^2}$. In (\ref{eq:ld}), ${\bf F}_{ext}$ is the external electromagnetic force and in this work we use ${\bf F}_{rad}=\Ld \tdot {\bf x}_e$, which is the lowest order approximation in powers of $(v/c)$ to the exact relativistic radiaton force\cite{Coleman}. For a circular orbit, this is a term of order $(v/c)^3$.  
\par
We now introduce the expansion of the retardation constraint of the Li\'enard-Wiechert interaction, henceforth called the Page series\cite{Page}. Let ${\bf x}_1 $ be the position of electron 1, of charge $-e$, and $\beta_1$ its velocity vector divided by $c$.
 The formula for the electric field made by electron 1 at a position ${\bf x}_2$ is
\begin{eqnarray}
{\bf E}_{12}&=&-e{ {\hat {\bf n}}\over{R_{12}^2}}+{\bf E}_{12}^{(2)}({\bf x}_2,{\bf x}_1,\dot {\bf x}_1, \ddot {\bf x}_1) + {\bf E}_{12}^{(3)}(\tdot {\bf x}_1)+\ldots
\label{TaylorE}
\end{eqnarray}
where
\begin{eqnarray}
{\bf E}_{12}^{(2)}\equiv -e\{ { {(|\beta_1|^2 -3(\hat {\bf n}\cdot \beta_1)^2)\hat {\bf n}}\over{2R_{12}^2} } -{{\dot \beta_1}\over{2cR_{12}}} -{{(\hat {\bf n} \cdot \dot \beta_1 )\hat {\bf n}}\over{2cR_{12}}} \}, \nonumber \\
\mbox{ }
\label{E212}
\end{eqnarray}
and
\begin{eqnarray}
{\bf E}_{12}^{(3)}(\tdot {\bf x}_1)\equiv -\Le \tdot {\bf x}_1,
\label{E312}
\end{eqnarray}
where $\hat {\bf n}$ is the unit vector in the direction of ${\bf x}_2-{\bf x}_1$ and $R_{12}=|{\bf x}_2-{\bf x}_1|$. Notice that all functions are evaluated at present time. For a circular orbit, in units of the zeroth order Coulomb term, (\ref{E212}) is of order $(v/c)^2$ and (\ref{E312}) is of order $(v/c)^3$. We assume that the Taylor series of (\ref{TaylorE}) converges, since $(v/c) \ll 1$ in atomic physics.   
The first term of the series for the magnetic field of electron 1 at ${\bf x}_2$ is\cite{Page}
\begin{eqnarray}
{\bf B}_{12}^{(1)}({\bf x}_2,{\bf x}_1,\dot {\bf x}_1)&=&-{{e}\over{R_{12}^2}}[\beta_1 \times \hat {\bf n}].
\label{SchotP-B}
\end{eqnarray}
In this work we consider only the above terms of the electromagnetic interaction, as well as the relativistic correction to the electronic dynamics up to order $(v/c)^2$.
\par 
Along a Wannier orbit, the total force acting on electron 2 can be calculated, using (\ref{E212}), (\ref{E312}), (\ref{SchotP-B}) and the self-interaction  of electron 2, to be 
\begin{eqnarray}
{\bf F}_2&=&\Ld (\tdot {\bf x}_1+ \tdot {\bf x}_2)-7{{e^2 \uni n}\over{4R^2}}(1-{{3}\over{7}}|\beta|^2),
\label{eq:LDDL}
\end{eqnarray}
where $R$ is the radius of the Wannier orbit. In (\ref{eq:LDDL}) we have also added the Coulomb attraction of the $\alpha$-particle (of charge $2e$), that we suppose infinitely massive and resting at the origin.  
Along the Wannier orbit, because the electrons are in phase opposition, the force along the velocity cancels out, demonstrating that the Wannier orbit is a possible periodic solution of the electromagnetic equations up to third order in $(v/c)$.
\par
Notice the appearance of the dipole term in (\ref{eq:LDDL}),
\begin{eqnarray}
\ddot {\bf D} &=& -e(\ddot {\bf x}_1 +\ddot{\bf x}_2 ).
\label{eq:dipole}
\end{eqnarray} 
The total far field caused by the electrons depends linearly on the quantity $\ddot {\bf D}$ of (\ref{eq:dipole}), up to quadrupole terms\cite{Jackson}. If ${\bf D}$ is zero, the orbit does not radiate in dipole. The fifth-order quadrupole terms would be important only in a much longer time scale, of order $T/(v/c)^5$\cite{deluca}. In this letter we consider only the dynamical system defined by truncating the Schott-Page interaction and self-interaction to third order in $(v/c)$. 
\par
Next we consider the dynamics of helium
in the neighborhood of a Wannier orbit, according to the Darwin Lagrangian. For the Coulomb dynamics, this stability problem has already been considered by
many authors\cite{Poirier,Nicholson}. Here we consider only the second order of the Page series, which is essential to unfold a degeneracy of the Coulomb dynamics. The detailed consideration of the dissipative effect of the third-order term of the Page series and the (also third-order) self-interaction term  will be published elsewhere\cite{deluca}. Here it suffices to say that these higher-order terms are important only in a much longer time-scale. To second order in $(v/c)$, the equations of motion for helium with retardation and relativistic correction are 
\begin{eqnarray}
M_e{{d}\over{dt}}[(1+{{1}\over{2}}|\dot {\bf x}_{1}|^2) \dot {\bf x}_1]=-{{2e^2}\over{R^3_{1}}}({\bf x}_{1}) -{{e^2}\over{R^3_{12}}}({\bf x}_{2}-{\bf x}_{1}) \nonumber \\ +{\bf E}_{21}^{(2)}({\bf x}_1,{\bf x}_2,\dot {\bf x}_2,\ddot {\bf x}_2)-{{e}\over{c}}\dot {\bf x}_1 \times {\bf B}^{(1)}_{21}({\bf x}_1,{\bf x}_2,\dot {\bf x}_2), \nonumber \\
M_e{{d}\over{dt}}[(1+{{1}\over{2}}|\dot {\bf x}_2|^2)\dot {\bf x}_{2}]=-{{2e^2}\over{R^3_{2}}}({\bf x}_{2}) -{{e^2}\over{R^3_{12}}}({\bf x}_{1}-{\bf x}_{2})\nonumber \\ +{\bf E}_{12}^{(2)}({\bf x}_2,{\bf x}_1,\dot {\bf x}_1,\ddot {\bf x}_1)-{{e}\over{c}}\dot {\bf x}_2 \times {\bf B}^{(1)}_{12}({\bf x}_2,{\bf x}_1,\dot {\bf x}_1),
\label{newtons}
\end{eqnarray}
where $M_e$ is the electronic mass, $R_{1}=|{\bf x}_1|$, $R_{2}=|{\bf x}_2|$, $R_{12}=|{\bf x}_1-{\bf x}_2|$, $ {\bf E}_{12}^{(2)}({\bf x}_2, {\bf x}_1,\dot {\bf x}_1,\ddot {\bf x}_1)$ is defined in (\ref{E212}) and $ {\bf B}^{(1)}_{12}({\bf x}_2,{\bf x}_1,\dot {\bf x}_1)$ is defined in (\ref{SchotP-B}). The Wannier orbit is the following periodic solution of (\ref{newtons}) 
\begin{eqnarray}
x_{1}=R\cos(\omega t), & y_{1}=R\sin(\omega t) & , z_{1}=0, \nonumber \\
x_{2}=-R\cos(\omega t), & y_{2}=-R\sin(\omega t), & z_{2}=0.
\label{rotsol}
\end{eqnarray}
According to (\ref{newtons}), the frequency of the orbit is related to $R$ by
\begin{eqnarray}
M_e \omega^2={{7e^2}\over{4R^3}}(1-{{3}\over{7}}|\beta|^2),
\label{Coulombdisper}
\end{eqnarray}
 which is correct to second order in $(v/c)$. 
\par 
Linearizing (\ref{newtons}) about the Wannier orbit (\ref{rotsol}), we obtain a parametric linear differential equation with coefficients periodic in time and period $T=\pi/\omega$\cite{deluca}. If the Floquet exponents are all nondegenerate, one can find a complete set of solutions of the form\cite{Jordan} 
\begin{eqnarray}
\delta {\bf x}_{1}&=&\exp(2i\omega \mu t)\sum_{n} {\bf x}^{1}_n\exp(2in\omega t),\
\nonumber   \\
\delta {\bf x}_{2}&=&\exp(2i\omega \mu t)\sum_{n} {\bf x}^{2}_n\exp(2in\omega t),
\label{Floquet}
\end{eqnarray}
where the Floquet exponent $\mu$ is a complex number defined in the first
Brillouin zone, $-1/2 <Re (\mu )< 1/2$. Notice that for the Floquet components we use an upper index, but to label coordinates as functions of time, as in (\ref{newtons}), we use a lower index (to distinguish it from the Floquet components).  
To bring the variational equations to normal form, we define the coordinates
\begin{eqnarray}
\xi^{\kappa}_n\equiv {{1}\over{\sqrt 2}}(x^{\kappa}_n-iy^{\kappa}_n), \mbox{  }\chi^{\kappa}_n \equiv {{1}\over{i\sqrt 2}} (x^{\kappa}_n+iy^{\kappa}_n). 
\label{defchi}
\end{eqnarray}  
Next we calculate Hill's secular determinant\cite{Jordan} for the linearization of (\ref{newtons}) about (\ref{rotsol}). As a simplification, let us define $\bar n \equiv n+\mu$ to be the running variable in the summations of (\ref{Floquet}). Last, to introduce physical intuition, it is convenient to define the radiation coordinate ${\bf x}_r$ and relative coordinate ${\bf x}_d$ as
\begin{eqnarray}
{\bf x}_{r}\equiv {\bf x}_{1}+{\bf x}_{2},\\
{\bf x}_{d}\equiv {\bf x}_{1}-{\bf x}_{2}.
\label{defradrel}
\end{eqnarray}  
It turns out, as one can easily verify, that the linearized dynamics for the $\delta {\bf x}_d$ variable is decoupled from the $\delta {\bf x}_{r}$ variation.
Here, for lack of space, we only develop the dynamics of the $\delta {\bf x}_d$ variable along the $z=0$ plane, correct to second order in $(v/c)$. We refer to \cite{Poirier} for the Coulomb eigenvalues of the $\delta {\bf x}_r$ dynamics. A complete study of the eigenvalues will be published elsewhere\cite{deluca}. 
Before we write the planar variational equations, we need still another definition
\begin{eqnarray}
U_n^{\kappa}\equiv{{1}\over{\sqrt 2}}\xi^{\kappa}_{n+1}+{{i}\over{\sqrt 2}}\chi_{n-1}^{\kappa}\nonumber \\
V_n^{\kappa}\equiv{{-i}\over{\sqrt 2}}\xi^{\kappa}_{n+1}-{{1}\over{\sqrt 2}}\chi_{n-1}^{\kappa}.
\label{defUL}
\end{eqnarray}
We make the quantities $U_n^{\kappa}$ and $V_n^{\kappa}$ into the x and y components of a bidimensional vector ${\bf K}_n^{\kappa}\equiv (U_n^{\kappa},V_n^{\kappa})$, and also define $\tilde {\bf K}_n^{\kappa}\equiv (-V_n^{\kappa},U_n^{\kappa})$, ${\bf x}_n \equiv (x_n,y_n)$ and $\tilde {\bf x}_n\equiv (-y_n^{\kappa},x_n^{\kappa})$. The variational equations for planar motion along the variable $\delta {\bf x}_d$ can be written most simply in Floquet components as\cite{deluca}
\begin{eqnarray}
-4\bar n^2 {\bf x}_n^{d} -{{1}\over{2}}({\bf x}_n^{d}+3{\bf K}_n^{d})\nonumber \\  +|\beta|^2(P(\bar n){\bf x}_n^{d}+Q(\bar n){\bf K}_n^{d} -{{6i\bar n}\over{7}}\tilde {\bf x}_n^{d} +{{18i\bar n}\over{7}}\tilde {\bf K}_n^{d} )=0.
\label{Wannierinst}
\end{eqnarray}
We recall that $\bar n=n+\mu$, and there is one equation for every value of $n$. In principle, this would lead to an infinite Hill's secular determinant\cite{Jordan}. For the case of circular orbits, when we use the $\xi$ and $\chi$ variables, we find that the $\xi_n$ variables couple only to $\chi_{n+1}$ and vice versa. The variational problem for the coordinate $\delta {\bf x}_d$, as defined by (\ref{Wannierinst}), turns into the following $2 \times 2$ matrix equation
\begin{eqnarray}
\left[ 
\begin{array}{cc}
-({{1}\over{2}}+4\bar n^2)&{{-3i}\over{2}} \\ 
{{3i}\over{2}}& -({{1}\over{2}}+4\bar n_{+}^2) 
\end{array}
\right] \left[ 
\begin{array}{c}
\xi^d_n \\ 
\chi^d_{n+1}
\end{array}
\right]\nonumber \\
+|\beta|^2 
\left[ \begin{array}{cc}
P_2(\bar n)& iQ_2(\bar n) \\ 
-iQ_2(-\bar n_{+}) & P_2(-\bar n_{+})  
\end{array}
\right] \left[ 
\begin{array}{c}
\xi^d_n \\ 
\chi^d_{n+1}
\end{array}
\right]=0.
\label{coulsecd}
\end{eqnarray}
where $P_2(\bar n)\equiv (-{{4\bar n^2}\over{7}}+{{6\bar n}\over{7}}+{{3}\over{28}})$, $Q_2(\bar n)\equiv ({{18\bar n^2}\over{7}}-{{18\bar n}\over{7}}+{{1}\over{28}})$ and $\bar n_{+}\equiv (1+\bar n)$. We now consider the secular equation associated with (\ref{coulsecd}): At $|\beta|=0$, we obtain the eigenvalues of the Coulomb stability problem
\begin{eqnarray}
\bar n=0,-1,-1/2,-1/2.
\label{roots}
\end{eqnarray}
The next step is to consider the secular equation of (\ref{coulsecd}) for finite values of $|\beta|$. The essential ingredient for us here is the degenerate root $\bar n=-1/2$ of (\ref{roots}). Because of this degeneracy, the double root $\bar n=-1/2$ acquires a correction proportional to $|\beta|$, instead of the generic second-order correction of the other nondegenerate roots. Solving the secular equation of (\ref{coulsecd}), in the neighborhood of $\bar n=-1/2$ and in the lowest order approximation in $|\beta|$, we find 
\begin{eqnarray}
\bar n \simeq -1/2 \pm\sqrt{{{51}\over{28}}} |\beta|.
\label{double}
\end{eqnarray}
According to (\ref{Floquet}), $\bar n$ is a frequency of oscillation in units of $2\omega$. For example, along the linear mode of (\ref{coulsecd}) defined by $\bar n=-1/2$, the frequencies in (\ref{Floquet}) are $-\omega=2\omega \bar n$ and $\omega=2\omega (1+\bar n)$. 
\par
Furta's theorem gives the condition for the existence of an analytic constant in the neighborhood of a fixed point of an autonomous vector field\cite{Furta}. The Wannier orbit is a fixed point of the autonomous vector field describing the dynamics in the system rotating with the frequency of the Wannier orbit. To apply the theorem, we must move to this rotating system. In this rotating system, the new frequency for oscillation along the mode (\ref{double}) is found by adding $1/2$ to (\ref{double}). In units of $2\omega$, the frequency of Coulombic oscillations along the $z$ direction in the mode $\delta z_r$ is found in \cite{Poirier} to be 
\begin{eqnarray}
\omega_1\equiv \sqrt{{2}\over{7}}\simeq 0.5345\ldots 
\label{freq27}
\end{eqnarray}
and for planar Coulombic oscillations along the mode $\delta {\bf x}_r$, in the rotating frame, is\cite{Poirier}
\begin{eqnarray}
\omega_2\equiv  \sqrt{{3+\sqrt 32}\over{28}}\simeq 0.5560\ldots
\label{posiWan}
\end{eqnarray}
also in units of $2\omega$. A new resonance condition involving the root (\ref{double}) is the easiest to be satisfied for small values of $|\beta|$. Therefore, we are looking for a resonance among the frequencies of (\ref{posiWan}), (\ref{freq27}), and (\ref{double}), which in the rotating frame evaluates to 
\begin{eqnarray}
\omega_3\simeq \sqrt{{51}\over{28}}|\beta| \simeq 1.3496 |\beta|,
\end{eqnarray}
in units of $2\omega$. According to Furta's theorem\cite{Furta}, for an additional analytic constant to exist one needs a new resonance among the linear frequencies. By inspection, we find that the resonance involving the above three frequencies and with the minimal integer multipliers is of type
\begin{eqnarray}
\omega_1-\omega_2+2n\omega_3=0,
\label{nres}
\end{eqnarray}
where $n$ is an integer number and the factor of two multiplying $\omega_3$ is because mode $3$ comes from a degenerate eigenvalue. 
Condition (\ref{nres}) is satisfied for $|\beta|$ given by
\begin{eqnarray}
|\beta|\simeq {{0.007969423}\over{n}}.
\end{eqnarray}
The binding energies, in atomic units, for the above values of $|\beta|$ are
\begin{eqnarray}
E=-{{2.3853}\over{n^2}} 
\end{eqnarray}
and the frequencies of the special ``stable'' Wannier orbits described by (\ref{nres}), in atomic units, are given by
\begin{eqnarray}
\omega={{0.7442}\over{n^3}}.
\end{eqnarray}
Because we are involving a resonance among the planar oscillations $\delta {\bf x}_r$, $\delta {\bf x}_d$, and the $\delta z_r$ oscillation, the leading term in the McLaurin series\cite{Furta} of the constant is a product of $\delta z_r$ times the planar coordinate variations. In reference\cite{deluca} we show how to construct this extra constant and discuss the shape of its level surfaces. 
\par
The frequency $\omega_z$ for oscillations along the $\delta z_{r}$ linear mode can be obtained, as prescribed by (\ref{freq27}), by multiplying the above frequency by $2\sqrt{2/7}$,
\begin{eqnarray}
\omega_z={{0.7956}\over{n^3}}.
\label{stabmani}
\end{eqnarray}  
	The inclusion of the third-order radiative terms makes this linear mode of (\ref{freq27}) a stable mode\cite{deluca}, whereby the perturbation radiates energy and decays to the perfectly circular orbit. The frequency of this stable oscillation should then appear
in the spectrum of sharp lines, given that the dynamics selects the neighborhood of the stable orbits. For example, the transition $2^{1}P\rightarrow 1^{1}S$ of parahelium corresponds to a frequency of $2.9037-2.1237=0.7799$ atomic units\cite{Bethe,Chen}, and (\ref{stabmani}) with $n=1$ evaluates to 0.7956, which is a $2\%$ difference. The transition $3^{1}P \rightarrow 2^{1}S$ in parahelium\cite{Chen} has a frequency of $2.1459-2.0551=0.0908$ atomic units, and (\ref{stabmani}) with $n=2$ evaluates to 0.0995, which is an $8\%$ difference. The asymptotic form of the quantum mechanical energies of helium are\cite{Bethe}
\begin{eqnarray}
E=-{{Z^2}\over{2}}-{{(Z-1)^2}\over{2(n_r+L+1)^2}},
\end{eqnarray}
with $Z=2$. For transitions with $\Delta L=1$, the frequencies predicted
from the above formula are given by
\begin{eqnarray}
\omega_q={{1}\over{(n_r+L+1)^3}},
\label{quantdiff}
\end{eqnarray}
which agrees with (\ref{stabmani}) to within $20\%$. We stress that we relied only on dynamics to calculate the frequency of (\ref{stabmani}). Equation (\ref{quantdiff}), on the other hand, was obtained in the usual way of taking differences between quantum energies, and was included only for comparison. 
\par
 A more detailed comparison with the spectrum of helium requires the consideration of elliptical orbits and zero-angular-momentum orbits\cite{deluca}. In reference \cite{deluca} we include the third-order terms and find that the resulting damping of the oscillation mode of (\ref{stabmani}) agrees well with the quantum calculations of the linewidth. 
\par
The extra complex constant is actually two real functions of phase space\cite{Furta,deluca}. Together with the four global constants of helium, there are six functions that can produce a barrier in phase space, at least in the small region inside the resonance islands\cite{Lichtenberg}. Projected on the plane of the stable and unstable Wannier directions, the one-period return map of the full dynamics has a hyperbolic fixed point. The resonance islands are possibly found inside the homoclinic tangle of the invariant manifolds, but this remains to be investigated in detail.    
\par 
Historically, the Wannier instability, first found by Nicholson\cite{Nicholson}, led Bohr to postulate that, along a quantized set of orbits, some special mechanism would supersede the Wannier instability, thus stabilizing the atom, which proved to be a very fruitful intuition\cite{Bohr}. Some of the details of how the six constants stabilize the dynamics, in a time scale of order $T/(v/c)^2$, are discussed elsewhere\cite{deluca}. 
\par
The existence of this extra constant is a genuine effect of electrodynamics with retardation. It is surprising that classical electrodynamics with retardation can offer such a sensible approximation to the spectrum of sharp lines. We do not know of any prior similar application of dynamics to atomic physics.
\par
I acknowledge discussions with R. Napolitano and J. C. Egues. This work was supported by Fapesp.

\end{document}